\magnification=\magstep 1
\tolerance=1600
\font\ner=cmbx10
\font\tit=cmr10 
\baselineskip= 0.85 true cm
\vsize= 20 true cm
\topinsert 
\vskip 50pt
\endinsert

{\ner
\centerline{Mobility Edge and Level Statistics}

\centerline{of Random Tight-Binding Hamiltonians}}

\vskip 15pt

{\tit
\hfill Fabio Siringo and Giovanni Piccitto \hfill}

\vskip 5pt

{\tit
\hfill 
Dipartimento di Fisica 
e Unit\`a INFM 
dell'Universit\`a di Catania \hfill}

{\tit
\hfill Corso Italia 57, I 95129 Catania - Italy\hfill}

\vskip 30pt

\centerline{\ner
Abstract }

\vskip 15pt

{\tit\narrower{
The energy level spacing distribution of a tight-binding hamiltonian
is monitored across the
mobility edge for a fixed disorder strength. 
Any mixing of extended and localized levels is avoided in
the configurational averages, thus approaching
the critical point very closely and with
high energy resolution. By finite size scaling
the method is shown to provide a very accurate estimate
of the mobility edge and of the critical exponent
for a cubic lattice with lorentzian distributed diagonal
disorder. Since no averaging in wide energy windows is required, the
method appears as a powerful tool for locating the mobility edges in
more complex models of real physical systems.
\smallskip}}

{\tit
}

\vfill\eject

{\tit
The Metal-Insulator (MI) transition in disordered systems is still
attracting considerable interest with special regard to the critical
universal properties [1,2]. 
Even in absence of any interaction, the lack of
a full analytical treatment for $d=3$ has given rise to a very slow
convergent process [3-6] for the numerical evaluations,
generally based on finite size scaling [7]. An alternative to
the standard transfer matrix method [8] has been recently given
by the Energy Level Statistics Method (ELSM) [9,10] which has reached the
status of a very reliable tool, yielding accurate predictions for the
transition points and the critical properties. Localized and delocalized
levels have been shown[9,10] to follow different universal spacing 
distributions $P(s)$ in the thermodynamic limit. Precisely, in the
metallic regime the overlapping states follow the general 
Wigner-Dyson random-matrix theory[11], and the distribution of the distances
between successive levels is well described by the Wigner surmise
$$P_W(s)={\pi\over 2} s \exp\left(-{\pi\over 4} s^2\right)\eqno(1)$$
which is characterized by the typical level repulsion since $P(0)=0$.
Here $s$ is the level distance in units of the average local level
spacing $\Delta E = 1/[ N D(E)]$ where $N$ is the total number of states
and $D(E)$ is the averaged Density of States (DoS) normalized to 1.
On the other hand for localized levels the lack of overlap determines
a different statistical behaviour described by the Poisson distribution
$$P_P(s)=e^{-s}\eqno(2)$$
where level repulsion is absent. For finite systems an intermediate
regime shows up in the cross-over region, and the corresponding
critical ensemble has been recently characterized and shown to be
described by a third universal distribution[12-14]. 
Then, in order to extrapolate towards the thermodynamic limit,
finite scaling has been
extensively used,  allowed by the
numerical calculation
of any one-parameter scaling function which characterizes
the distribution $P(s)$ along the cross-over[4,13].
The critical ensemble which determines the critical properties has been
recently shown to be universal and even insensitive to the changes 
of symmetry determined by a field[15] or spin-orbit coupling[16], 
or to changes
in the nature of the disorder[17]. 
Thus the method seems to be even of practical
interest for determining the critical properties of more realistic 
phenomenological models.

However, although large clusters have been recently considered up to
$28^3$ sites [6], some shortcomings can be found in the generally
accepted ELSM procedure: i) In order to improve the statistical
analysis a wide energy band is usually considered, thus averaging over
a dishomogeneous class of levels. In fact, at the critical point,
the mixing of extended and localized levels cannot be ruled out even
for a box distribution which is known to yield a very flat behaviour.
As a consequence the critical point cannot be reached with good
accuracy.
ii) Only a box distribution for the diagonal matrix levels  is well
suited for such approach, since for a generic peaked distribution 
the use of a wide band would be unacceptable. Thus any realistic
calculation, based on a phenomenological model, is ruled out.
iii) The scaling function is usually evaluated for discrete fixed values
of the disorder strength parameter $W$, and the critical point is then
recovered by interpolation. Since any new value of $W$ requires a new
average over the configurations, a very small step for $W$ around the
critical point becomes prohibitive.
iv) No information is obtained about the trajectory of the mobility
edge versus $W$ below the critical point.

In this paper, in order to deal with such difficulties, we propose the
first attempt to characterize the transition through a scaling function
of the level energy $E$. Below the critical disorder, a mobility edge
separates localized and extended levels in the thermodynamic limit. 
Thus the spacing distribution $P(s)$ changes discontinuously at the
mobility edge from $P_P(s)$ to $P_W(s)$. For a finite system the transition
is rounded and a continuous cross-over is observed. Any monotonic
functional $\alpha$ of $P(s)$ could in principle be employed for 
characterizing the spacing distribution as a function of the energy $E$
across the transition[4].
According to the one-parameter scaling theory of localization [18]
the coherence length of the states is a function of the level energy
$\xi (E)$ diverging at the mobility edge $E_c$ as
$$\xi\sim (E-E_c)^{-\nu}\eqno(3)$$
For a finite system, neglecting boundary effects, 
the functional $\alpha$ depends on $E$ through the
only parameter $\xi(E)/L$ where $L$ is the linear dimension of the system.
Thus finite size scaling yields the mobility edge $E_c$ and the critical
exponent $\nu$ provided that the functional $\alpha$ can be determined as a
function of $E$ across the transition, for a fixed disorder strength $W$.
We choose for $\alpha$
the second moment $\alpha=\langle s^2\rangle$, 
and directly
evaluate such functional on the level configurations for any $E$ by a new
method which avoids both, averaging
inside a wide window of levels, and any fit for the distribution $P(s)$.
Apart from statistical fluctuations, we obtain $\alpha$ as a 
continuous function of $E$ across the mobility edge which can be approached
with the desired accuracy within the same set of configurations (only
one set of configurations, with a fixed disorder distribution, 
is required for each value of $L$). Moreover we avoid any mixing between
localized and extended states in the evaluation of $\alpha$ which now 
acquires a different value for any different energy. 

In order to illustrate the method we consider a standard tight-binding
hamiltonian on a cubic lattice with nearest-neighbour hopping and 
diagonal disorder described by a peaked lorentzian 
$$g(\epsilon)={1\over \pi}{W\over{W^2+\epsilon^2}}\eqno(4)$$
The mobility edge is obtained for several values of $W$, and the critical
exponent $\nu=1.30\pm0.05$ is recovered in excellent agreement with
previous reports[3,4,13,16]. We notice that the fieldtheoretical argument
reported in Ref.[3] for the universality of the critical exponent
does not hold for the lorentzian distribution since now the moments are
not well defined. Thus the agreement with the estimates obtained for
the box and gaussian distribution[3], and for the binary percolation
model[17] should be understood in the framework of a more general 
theoretical argument.

The hamiltonian reads
$$\hat H=\sum_{i} \epsilon_i c_i^\dagger c_i+t\sum_{\langle ij\rangle}
c_i^\dagger c_j\eqno(5)$$
where $c_i$, $c_i^\dagger$ are annihilation and creation 
operators for a local state on the cubic lattice site $i$, and the
diagonal level $\epsilon_i$ is randomly distributed according to the
lorentzian (4). All energies are reported in units of the 
nearest-neighbour hopping term $t$, thus the system is entirely 
characterized by the width $W$ of the distribution.

The eigenvalues are exactly evaluated for each random
configuration of a finite $M\times M\times M$ cubic lattice with
$M=6,7,8,9,10$. No special boundary conditions have been imposed 
and a total number of $10^4$ configurations have been
considered for each lattice size $M$.

The choice of the second moment $\alpha=\langle s^2\rangle$ for 
characterizing the spacing distribution comes from statistical
convenience: no fit is required or knowledge of the distribution, since
$\alpha$ is computed by direct average over the configurations. Moreover
all the configurations contribute in the average, and both the large $s$
and small $s$ regions are taken in due account. Such functional has a
monotonic behaviour from the Poisson limit value $\alpha=2$ to the
Wigner surmise opposite limit $\alpha={4 / \pi}\approx 1.27$, as
can be easily checked from equation (1) and (2). 

The second moment has already been used for characterizing the spacing
distribution $P(s)$ as a function of the strength $W$ of the disorder [13].
In that work a wide energy window was used in order to decrease the
statistical fluctuations. Conversely in the present work, 
at the cost of a larger fluctuation
of $\alpha$, we evaluate a basically independent average for each value
of $E$,  at a fixed disorder strength. For any energy $E$, and for each
configuration, we find two consecutive eigenvalues $E_i$, $E_{i+1}$ 
satisfying $E_i<E<E_{i+1}$. Then the variable $y=(E_{i+1}-E_i)/\Delta E$ is
averaged over the $N_{conf}$ random configurations of the system. Notice
that the average level spacing $\Delta E$ is a function of $E$, and is
independently evaluated by the average DoS. Since each couple 
of consecutive
eigenvalues has a probability proportional to $y$ of containing the point
$E$, then the generic moment follows
$$\langle s^n \rangle=\langle y^{n-1} \rangle \eqno(6)$$

From the central limit
theorem the statistical fluctuation of $\alpha$ may be estimated as
$\Delta \alpha^2\approx 
(\langle s^4 \rangle-\langle s^2\rangle^2 )/N_{conf}$.
In the Poisson limit, where the fluctuations are larger, 
and for $N_{conf}=10^4$, we obtain from eq.(2)
$\Delta\alpha\approx 0.045$ which is slightly more than 2 per cent.

It is instructive recovering the Poisson limit for $t=0$, and for a generic
distribution $g(\epsilon)$. Since the hamiltonian is diagonal, the
DoS is directly given by the distribution $g$ for the diagonal level.
At a fixed energy $E$ the average spacing is $\Delta E=1/[Ng(E)]$ where
the total number of states is $N=M^3$. Following an argument first
considered by Hertz[19], the probability $P(s) ds$ must be equal to
the probability that no levels exist between $E$ and $E+s\Delta E$ 
times the probability that a level does exist between $E+s\Delta E$ and
$E+(s+ds)\Delta E$:
$$P(s)ds=\left[1-\int_0^s P(s^\prime)d s^\prime\right] 
N\Delta E g(E+s\Delta E)ds\eqno(7)$$
Dividing by $g(E+s\Delta E)ds$ and differentiating with respect to $s$
we obtain
$${{P^\prime (s)}\over{P(s)}}=-{1\over{g(E+s\Delta E)}}\left[
g(E+s\Delta E) -{{g^\prime (E+s\Delta E)}\over{N g(E+s\Delta E)}}\right]
\eqno(8)$$
Such equation may be directly integrated, yielding a different distribution
$P(s)$ for each value of $E$. In the very special case of a box 
distribution the function $g$ is constant and the Poisson distribution
is recovered for any $E$. In general, assuming $N$ large, we may expand
the right hand side of equation (8) in powers of $s\Delta E$, then
integrating
$$P(s)=P(0)\exp\left[-s-{{g^\prime}\over{Ng^2}}({1\over 2} s^2-s)+0(1/{N^2})
\right]\eqno(9)$$
Thus, for large $N$, the distribution $P(s)$ tends to the Poisson universal
limit at any $E$, and for any distribution $g$. However the convergence is
not uniform, since it is controlled by the parameter $g^\prime/(Ng^2)$ which
is strongly energy dependent and diverging in the limit $E\to\infty$ for
any regular normalized distribution. In other words, deep in the tail of
the DoS a larger and larger system size is required for recovering the
Poisson limit $\gamma\to 2$ as the energy $E$ increases. Strong deviations
from the limit distribution are then expected in the tails, where 
the size required for the convergence could be prohibitive for any
numerical calculation. Once more we show the importance of avoiding
level mixing inside a wide energy window, since not all the levels are
generally described by the same spacing distribution even in the case of
strong localization ($t=0$). The convergence towards the Poisson limit
must be checked as a function of $E$ before undertaking any serious
calculation, in order to fix a bound to the accessible range of energy
for a given system size. For the lorentzian (4) 
$g^\prime/(Ng^2)\sim E/(NW)$ and, as we checked by a numerical test, for
$N\approx 10^2-10^3$, up to $E\approx 7W$ the deviations from the Poisson
limit $\alpha=2$ are negligible as shown in figure 1. In such figure
the functional $\alpha$ is compared for $t=0$ and $t\not=0$, at
a fixed disorder strength and sample size.
For $t\not=0$ the second moment $\alpha$ approaches 
the Wigner limit around the band center, indicating that for such energy
range the states are extended. In the band tail $\alpha$ tends to the
Poisson limit, and a cross-over region shows up around $E=E_c$.

In order to get rid of the statistical fluctuations of $\alpha$, the data
around $E_c$ have been fitted by a polynomial. 
Data for $W=2$ are shown in figure 2: 
for different sizes $L=M-1$  
the fitted curves all cross at one point 
with surprising
accuracy: in fact the fit procedure allows an averaging over a large
number of fluctuating estimates of $\alpha$, thus improving the
accuracy without spoiling the energy dependence of the curves.
The crossing point
$E=E_c$ corresponds to the
mobility edge of the infinite system. For $W=2$ averaging over all the
crossing points yields the very accurate
determination of the mobility edge $E_c=3.8\pm 0.5$. 

According to the one-parameter scaling hypothesis, linearizing $\alpha$
around $E_c$, we obtain
$$\log L=\nu \log\vert\alpha^\prime(E_c,L)\vert+const\eqno(10)$$
where the derivative $\alpha^\prime$ can be evaluated with good accuracy
at the critical point by the polynomial fit parameters. Thus the method
allows a very close approach to the critical point, while avoiding any
mixing between localized and extended states. The critical exponent
$\nu$ has been determined 
by a linear fit of equation (10) for 
$M=6,7,8,9,10$ yielding $\nu=1.35\pm 0.05$ for $W=2$ and $\nu=1.26\pm 0.05$
for $W=2.5$.
The average value $\nu=1.30$
is very close to previous estimates obtained by several
authors for the box distribution[3,4,9,13], for the gaussian and
binary distributions[3], in presence of a field[15] or of spin-orbit
coupling[16], and even for the off-diagonal binary disorder 
of a percolation system[17]. All these estimates were found by 
averaging over a more or less wide energy window of levels at the center
of the DoS. In such case the monitored transition corresponds to the
disappearance of any extended state from the system. For the percolation
system the occurrence of a more realistic peaked DoS has forced the
authors[17] to focus on narrow windows. A more detailed study of the
mobility edge in that system would be achievable by the present method.
All these predictions are in slight contrast to recent findings by 
transfer matrix method [5]
and standard ELSM[6] reporting a larger estimate for the critical exponent
($\nu=1.45$ and $\nu=1.54$ respectively) which is not confirmed by the
present work.

A tentative trajectory for the mobility edge has been recovered by
repeating the calculation at different values of $W$. No crossing
occurs for $W>4$, indicating that the critical point must lie below
that value. For $W=1.5$, $2.0$, $2.5$, $3.0$, $3.25$ and $3.5$
the mobility edge has been found
at the energy $E_c=4.2$, $3.8$, $3.3$, $2.4$, $1.8$ and $0.0$ respectively. 
Thus we may locate the critical disorder at $W_c=3.5\pm 0.1$.

As an important by-product, we have noticed the existence of a
universal critical value $\alpha_c=1.5\pm 0.03$
for all the considered disorder strengths $W$, thus
pointing out towards the existence of a critical $P(s)$ distribution
which is expected to be universal[12-14]. Our estimate for $\alpha_c$ is
not too far from the previous evaluation $\alpha_c=1/0.7=1.43$
reported by Varga et al.[13] for a box distribution. 

In summary we have shown that the ELSM provides a very accurate way of
determining the mobility edge and the critical exponent even for more
realistic disorder distributions. The scaling function has been determined
for any energy point $E$ avoiding any mixing of localized and 
extended states in the configurational average. 
Moreover a standard fit procedure allows
a very close approach to the critical point with a small statistical
error without spoiling the energy dependence of the scaling function.
Thus the method provides a powerful tool for detecting the mobility
edge in the framework of more complex phenomenological models of
real physical systems. As a by-product we have shown that the lorentz
distributed random system belongs to the same universality class of
the box and the gaussian, as evident from a comparison of the critical
exponents. We notice that, in spite of being generally expected,
such universality cannot be proved by the weaker fieldtheoretical
argument of Ref.[3] which is based on the assumption that the 
distribution has well defined moments.
\vfill\eject

{\ner
REFERENCES}

\vskip 15pt

{\tit
\item {[1]} P.A.Lee, T.V.Ramakrishnan, {\it Rev.Mod.Phys.} {\bf 57}, 
287 (1985).
\item {[2]} B.Kramer, A.MacKinnon, {\it Rep.Prog.Phys.} {\bf 56},
1469 (1993).
\item {[3]} E.Hofstetter, M.Schreiber, {\it Europhys.Lett} {\bf 21},
933 (1993).
\item {[4]} E.Hofstetter, M.Schreiber, {\it Phys.Rev.B} {\bf 49},
14726 (1994).
\item {[5]} A.MacKinnon, {\it J.Phys.:Condens.Matter} {\bf 6}, 
2511 (1994).
\item {[6]} I.K.Zharekeshev, B.Kramer, {\it Phys.Rev.B} {\bf 51},
17239 (1995).
\item {[7]} K.Binder, {\it Phys.Rev.Lett.} {\bf 47}, 693 (1981).
\item{[8]} A.MacKinnon, B.Kramer, {\it phys.Rev.Lett} {\bf 47}, 
1546 (1981).
\item{[9]} B.I.Shklovskii, B.Shapiro, {\it Phys.Rev.B} {\bf 47},
11487 (1993).
\item{[10]} E.Hofstetter, M.Schreiber, {\it Phys.Rev.B} {\bf 48},
16979 (1993).
\item{[11]} E.P.Wigner, {\it Ann.Math.} {\bf 62}, 548 (1955); {\bf 65},
203, (1957). F.J.Dyson, {\it J.Math.Phys.} {\bf 3}, 140 (1962); {\bf 3},
1199 (1962).
\item{[12]} V.E.Kravtsov, I.V.Lerner, B.L.Altshuler, A.G.Aronov,
{\it Phys.Rev.Lett.} {\bf 72}, 888 (1994); A.G.Aronov, V.E.Kravtsov,
I.Lerner, {\it Phys.Rev.Lett.} {\bf 74}, 1174 (1995).
\item{[13]} I.Varga, E.Hofstetter, M.Schreiber, J.Pipek, {\it Phys.Rev.B}
{\bf 52}, 7783 (1995).
\item{[14]} S.N.Evangelou, {\it Phys.Rev.B} {\bf 49}, 16805 (1994).
\item{[15]} E.Hofstetter, M.Schreiber, {\it Phys.Rev.Lett.} {\bf 73},
3137 (1994).
\item{[16]} T.Kawarabayashi, T.Ohtsuki, K.Slevin, Y.Ono, 
{\it Phys.Rev.Lett.} {\bf 77}, 3593 (1996).
\item{[17]} R.Berkovits, Y.Avishai, {\it Phys.Rev.B} {\bf 53}, 
R16125 (1996).
\item{[18]} E.Abrahams, P.W.Anderson, D.C.Licciardello, T.V.Ramakrishnan,
{\it Phys.Rev.Lett.} {\bf 42}, 673 (1979).
\item{[19]} P.Hertz, {\it Math. Ann.} {\bf 67}, 387 (1909).
\item{}
}

{\ner
Figure Captions}

\vskip 15pt

{\tit
\item{Fig.1} The second moment $\alpha$ versus the energy E inside the
band for $t=0$ (full circles) and $t\not=0$ (open circles). The disorder
strength is fixed at $W=2$, and the sample size is $M=10$. 
All the energies are in units of $t$.
An arrow in the cross-over region shows the position of the mobility edge 
$E_c=3.8$ as determined by finite size scaling. The dashed line is the
corresponding averaged DoS scaled by an arbitrary factor.
\item{Fig.2} Polynomial fit of the data for $\alpha$
in the cross-over energy range, for $M=6,7,8,9,10$ and $W=2$. At the band
center $E\approx 0$ the lower curve corresponds to the largest size.
\item{}
}

\vfill\eject

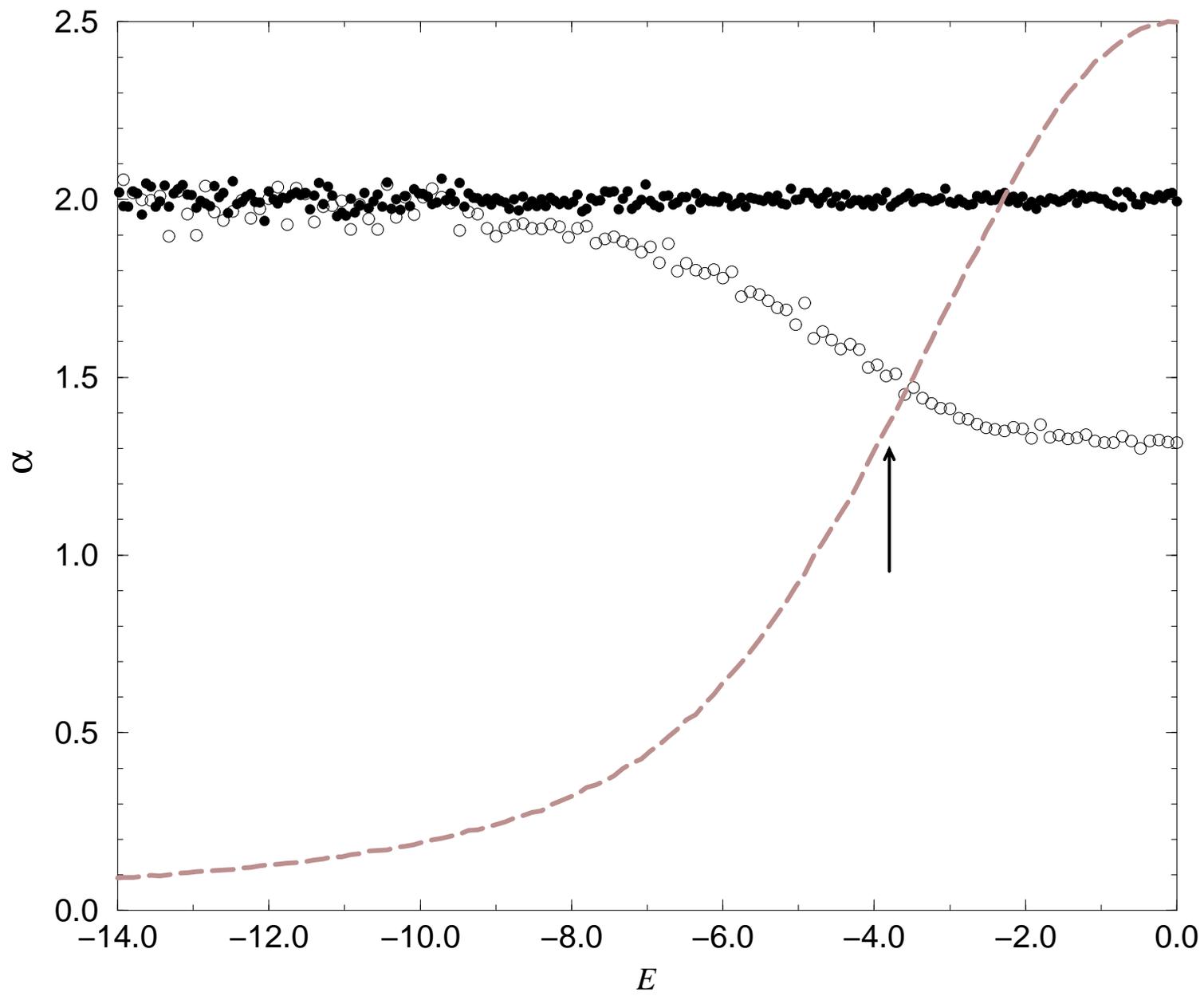
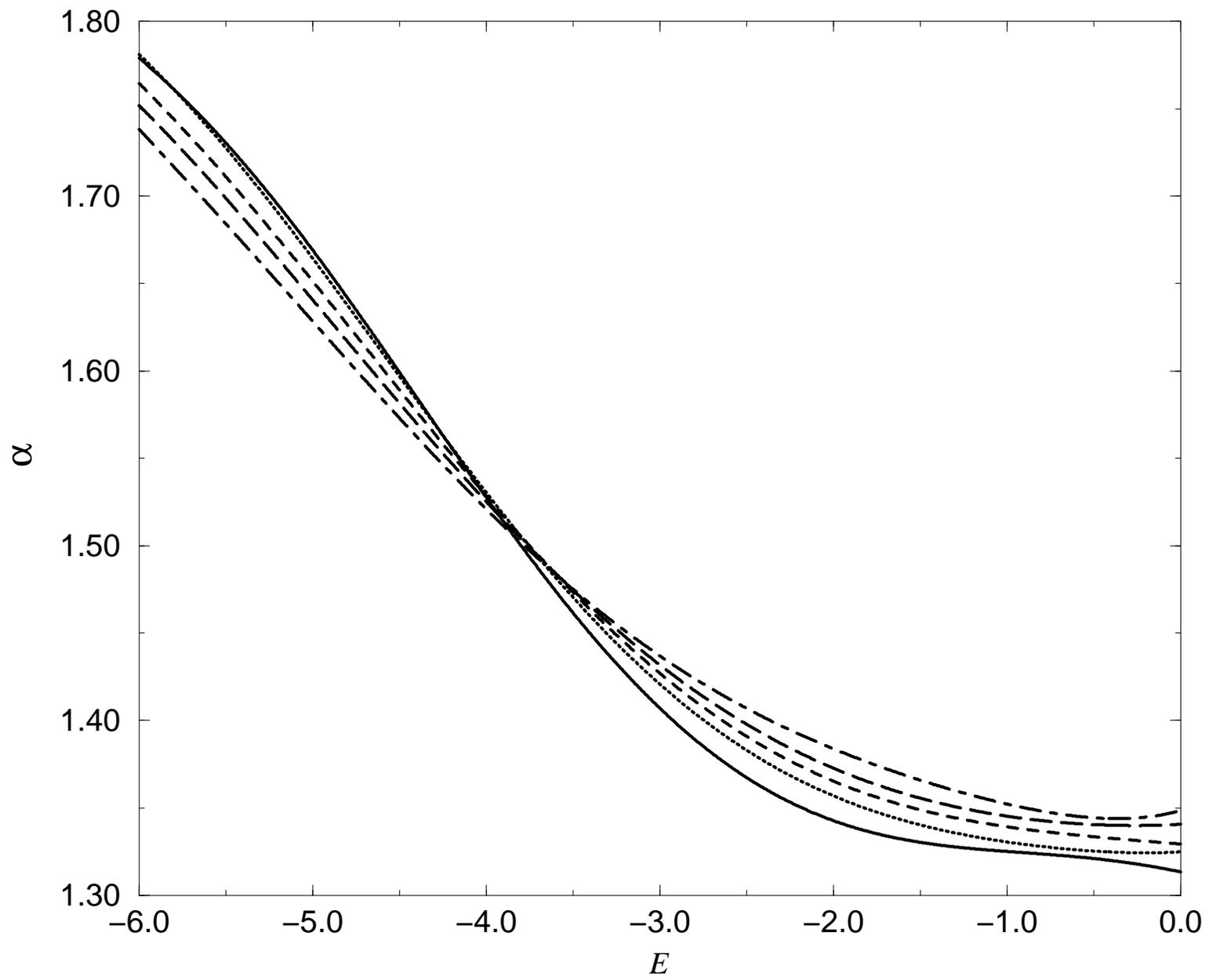
\bye